\documentclass[english,preprint,11pt]{elsarticle}
\pdfoutput=1
\usepackage[T1]{fontenc}
\usepackage[latin9]{inputenc}
\usepackage{geometry}
\usepackage{array}
\usepackage{float}
\usepackage{units}
\usepackage{textcomp}
\usepackage{multirow}
\usepackage{amsmath}
\usepackage{graphicx}
\usepackage{etoolbox}

\makeatletter
\patchcmd{\ps@pprintTitle}{Preprint submitted to}{}{}{}
\patchcmd{\ps@pprintTitle}{Elsevier}{}{}{}
\makeatother

\makeatletter

\providecommand{\tabularnewline}{\\}

\def\itoslometaddress{%
  Department of Information Technology, %
  Oslo Metropolitan University. %
}
\def\itoslometsimulametaddress{%
  Department of Information Technology, %
  Oslo Metropolitan University and 
  Simula Metropolitan Center.
}
\def\mekoslometaddress{%
  Department of Mechanical, Electronics and Chemical Engineering, %
  Oslo Metropolitan University. %
}
\def\hsoslometaddress{%
  Department of Health Science, %
  Oslo Metropolitan University. %
}
\def\uspmataddress{%
  Institute of Mathematical Science and Computing, %
  University of S\~ao Paulo, S\~ao Carlos.%
}

\author[mek]{Marco A. Pinto-Orellana}%
\author[usp]{Diego C. Nascimento}%
\author[mek]{Peyman Mirtaheri}%
\author[hs]{Rune Jonassen}%
\author[it]{Anis Yazidi}%
\author[its]{Hugo L. Hammer}%

\address[mek]{\mekoslometaddress{}}%
\address[usp]{\uspmataddress{}}%
\address[hs]{\hsoslometaddress{}}%
\address[it]{\itoslometaddress{}}%
\address[its]{\itoslometsimulametaddress{}}%

\makeatother

\usepackage{babel}
\begin{document}

\title{A hemodynamic decomposition model for detecting cognitive load using functional near-infrared spectroscopy}
\begin{abstract}
In the current paper, we introduce a parametric data-driven model for functional near-infrared spectroscopy that decomposes a signal into a series of independent, rescaled, time-shifted, hemodynamic basis functions. Each decomposed waveform retains relevant biological information about the expected hemodynamic behavior. The model is also presented along with an efficient iterative estimation method to improve the computational speed. Our hemodynamic decomposition model (HDM) extends the canonical model for instances when a) the external stimuli are unknown, or b) when the assumption of a direct relationship between the experimental stimuli and the hemodynamic responses cannot hold. We also argue that the proposed approach can be potentially adopted as a feature transformation method for machine learning purposes. By virtue of applying our devised HDM to a cognitive load classification task on fNIRS signals, we have achieved an accuracy of 86.20\%\textpm 2.56\% using six channels in the frontal cortex, and 86.34\%\textpm 2.81\% utilizing only the AFpz channel also located in the frontal area. In comparison, state-of-the-art time-spectral transformations only yield 64.61\%\textpm 3.03\% and 37.8\%\textpm 2.96\% under identical experimental settings.
\end{abstract}

\maketitle

\section{Introduction}

Functional near-infrared spectroscopy is a noninvasive neuroimaging technique that measures the hemodynamic response of the brain based on the change of light absorption in the near-infrared spectrum (700nm--900 nm) \cite{kim2017application}. Light is emitted on the scalp, and crosses the skull, cerebrospinal fluid, and reaches the cerebral cortex, where the hemodynamic chromophore components scatter and absorb the light according to their optical properties \cite{herold2018applications}. The primary hemodynamic components in the blood are mainly oxy-hemoglobin (HbO) and deoxy-hemoglobin (HbR) \cite{Tak_2014_C4}. The modified Beer-Lambert law provides a non-linear relationship that associates fNIRS light intensities with estimates of the local concentration of those components. Based on the neurovascular coupling theory, the information about the oxygenation changes are based on metabolic demands during the brain neural activations \cite{Herrmann_2017_C1}. Therefore, fNIRS can be seen as an important brain activity  monitoring technique that is complementary to other techniques such as electroencephalography which reflects only the electrical activation of the neurons \cite{Pellegrino_2016_C3}.

Optical spectroscopy, as a brain activity imaging method, is relatively new in the neuroscience community. However, the underlying process measured by the latter technique: either metabolic or hemodynamic responses of the brain, has been already widely studied through blood-oxygen-level-dependent signal (BOLD) in functional magnetic resonance imaging (fMRI) over the last three decades \cite{Y_cel_2016_F6}. Even though fMRI studies are clinically accepted as accurate to identify brain physiological activities, the inherent cost of the equipment, the lack of portability, and the low time resolution, put a restriction on the number and type of experiments that can be performed and analyzed. In contrast, fNIRS provides a reliable alternative to collect hemodynamic data with a time resolution that can be up to ten times higher than their fMRI counterpart, with enhanced portability. However, this comes with the cost of a lower spatial resolution \cite{Hennrich_2015_B}.

The research in the field of fNIRS falls into two main families according to their respective focus: clinical and classification oriented studies. Those two research communities aspire to investigate the differences in the signal under the effect of stimuli with their own custom approaches. Clinical studies often focus on quantifying the variations in the recordings when a particular stimulus is applied \cite{herold2018applications}. On the other hand, classification-oriented studies center their goal on the discovery of nonlinear relationships between stimuli and optical signals in order to devise models to predict the provoking stimulus from a new set of signals, with a certain degree of accuracy \cite{Moghimi_2012_D}.

In clinical settings, fNIRS studies rely on the comparison of time-domain properties of the signals. It is assumed that the underlying biological process is a linear time-invariant stimulus-guided process with a noise component, which is modeled as a Gaussian random variable \cite{Tak_2014_C4}. Hence, the signal amplitude difference between a phenomenon of interest and a control case along with its statistical significance can be assessed through the estimation of the first and second statistical noncentral moments. The uncertainty of this difference is usually expressed through confidence intervals or p-values \cite{Herrmann_2017_C1,Pellegrino_2016_C3,Wiggins_2016_C2}.

Clinical procedures are shown to be accurate in revealing specific spatial-temporal patterns in brain activity. It was determined that particular regions of the brain are more \textquotedbl activated\textquotedbl{} than others under specific circumstances or stimuli. It should be perceived that, in this context, \textquotedbl activation\textquotedbl{} denotes a high instantaneous amplitude with respect to the baseline observed mean. 

A common method in experimental BOLD-fMRI studies is modeling the fMRI signal with a generalized linear model (GLM) where the design matrix - i.e., the set of regressors of interest - is constructed with a sequence of canonical hemodynamic response functions (HRFs) \cite{Mumford_2015_F1}. HRFs are significant models of time-domain brain metabolic activity, providing a series of measures with psycho-physiological interest: amplitude, delay, and duration of the hemodynamic process \cite{Handwerker_2004_F3}. GLM-HRF in fNIRS data is known to be reliable for detecting differences in the activation patterns in the visual cortex \cite{Huppert_2016, Y_cel_2016_F6}.

For classification-oriented tasks, machine learning methods exhibit high performance in activity prediction due to their ability to explore and discover nonlinear structures in the data. In contrast with other classification fields, recent studies have been inconclusive in determining a predominant algorithm that can outperform all classification tasks when fNIRS is involved. Levels of pain can be detected with support vector machines with 94\% accuracy \cite{Fernandez_Rojas_2019_A}. Differentiation between word generation, motor imagery, and mental arithmetic yielded an accuracy as low as 66\% with a deep neural network \cite{Hennrich_2015_B} while discrimination between visual and auditory stimuli reached an accuracy of 97.8\% with support vector machine (and additional features extracted from EEG) \cite{Putze_2014_C}. In contrast to the nonlinear models, the accomplished accuracy in emotion classification was 71.9\% using linear discriminant analysis (LDA) \cite{Moghimi_2012_D}. Moreover, speech activity identification obtained an accuracy of 74.7\% using generalized canonical correlation analysis \cite{Liu_2018_E}. This diversity of algorithms with a large variability of accuracy coincides with the no free lunch theorem that states that \textquotedbl \emph{no algorithm can be good for any arbitrary prior}\textquotedbl{} information \cite{Adam_2019_NFL}.

Despite the effectiveness of machine learning methods in classification oriented studies, the complexity associated with the interpretability of the nonlinear models, generally more flexible, although limits their use in clinical studies \cite{Fernandez_Rojas_2019_A}. In the latter, the explanation of underlying physiological processes is more valuable \cite{Marrelec_2004_H6}. Mathematical models with enough complexity to provide high predictive performance, but based on interpretable and physiological-supported models, then this gap shall be solved.

In this paper, we propose a parametric method: the hemodynamic decomposition model (HDM), which aims to satisfy the requirements of both types of fNIRS analysis. HDM explains a metabolic signal, such as oxy- or deoxy-hemoglobin, as a sum of a series of independent bases. Each basis function is a simplified version of the canonical hemodynamic response function for sensorimotor activities while keeping a clear theoretical description of the underlying metabolic activity. HDM also provides enough information as a feature input for prediction-oriented purposes showing an exceptional predictive power in machine learning algorithms under different circumstances. 

The remainder of this paper is organized into three additional sections: the description of the model and the iterative estimation method (section 2); the analysis of the potential of HDM for machine learning tasks using real fNIRS data (section 3); and our conclusions (section 4.)

\section{Model}
\begin{center}
\begin{figure*}[th]
\begin{centering}
\includegraphics[width=0.95\textwidth]{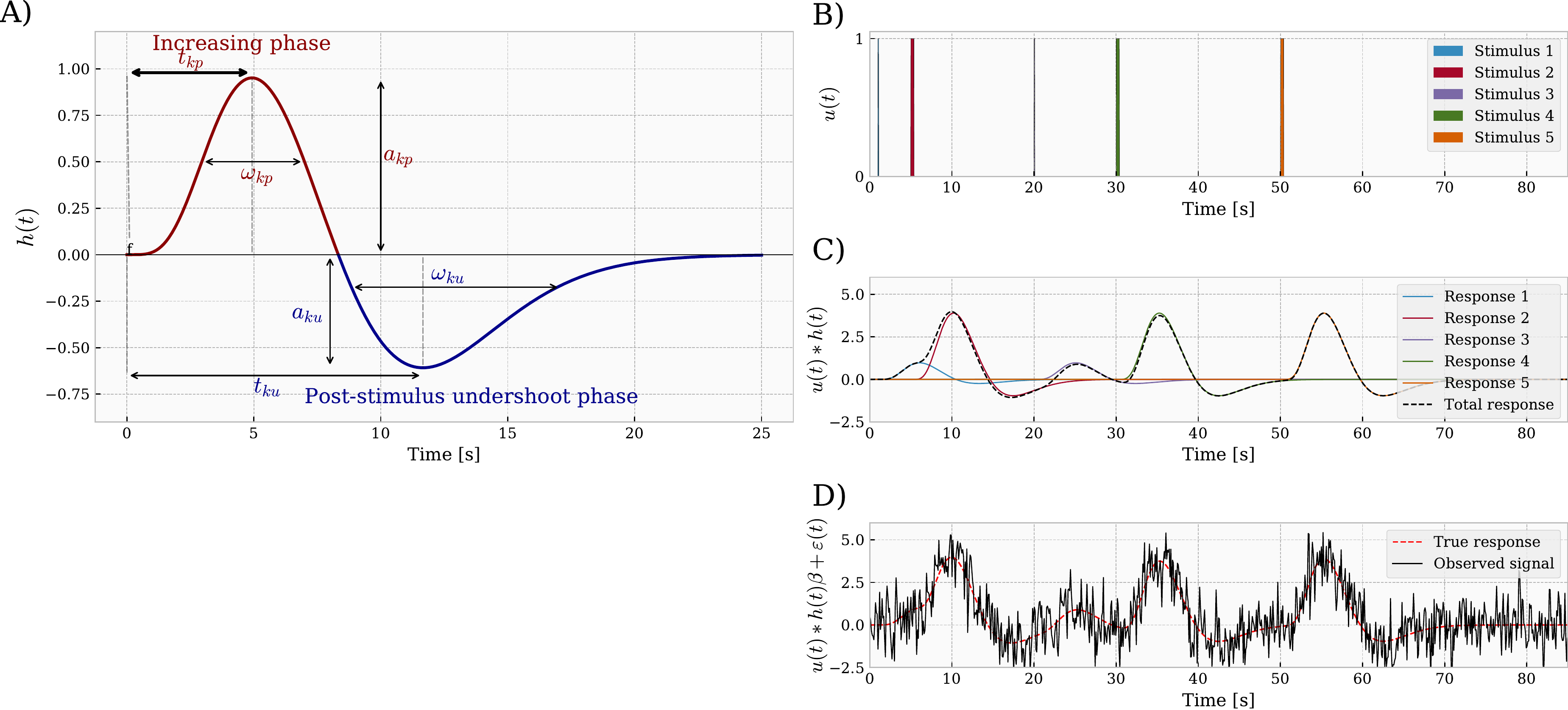}
\par\end{centering}
\caption{\label{fig:Simulation-GLM}Subject-level GLM of a hemodynamic data with the canonical HRF. (A) Components of the canonical double-gamma HRF, considering the GLM. (B) Hypothetical triggered set of stimuli $u\left(t\right)$. (C) Hemodynamics modelling, estimating the causes a change, with a convolution operation $u\left(t\right)\ast h\left(t\right)$. (D) Empirical distorted and noised signal $u\left(t\right)\ast h\left(t\right)+\varepsilon\left(t\right)$. These four panels represent the canonical HRF decomposition for sensorimotor activities.}
\end{figure*}
\par\end{center}

Hemodynamic signals, within-subjects, are modeled as linear time-invariant systems using theoretical hemodynamic response functions (HRFs) as a foundation. An HRF is a function that represents the typical oscillation phases in the blood dynamics. The de facto model for hemodynamic data, originally defined for fMRI time series, defines the output signal $y\left(t\right)$ as the convolution of a stimulus function $u\left(t\right)$ and an HRF $h\left(t\right)$ with an additive noise component \cite{Cignetti_2016_H0,Friston_1998_H3}: 
\begin{align}
y\left(t\right) & =x\left(t\right)\beta+\varepsilon\left(t\right)\\
x\left(t\right) & =u\left(t\right)\ast h\left(t\right)
\end{align}
where $\ast$ is the convolution operator, $\beta$ is the regression parameter, while $\varepsilon\left(t\right)$ models the noise. Subsequent measurements in BOLD-fMRI signals show correlated noise. Therefore, this behavior is modeled through a first-order autoregressive noise \cite{Seghouane_2015_H8}: 
\begin{equation}
~\varepsilon\left(t\right)=\rho\varepsilon\left(t-1\right)+\eta\left(t\right)
\end{equation}
where $\eta\left(t\right)$ is a Gaussian white noise $\eta\left(t\right)\sim\mathcal{N}\left(0,\sigma^{2}\right)$ with a variance $\sigma^{2}$. A simulation of the hemodynamic representation performed by this model is depicted in Figure \ref{fig:Simulation-GLM}b-d.

It is worth mentioning that there is not a unique and general HRF, and in order to guarantee that the GLM's parameters could be estimated from the data, there are three conditions that a HRF must meet: a) to start and end at zero; b) to be smooth, or at least, have continuity in the first and second derivatives \cite{Marrelec_2004_H6}; and c) to be a causal function, i.e., it should only depend on current or previous values \cite{Makni_H7}.

Several types of basis HRFs that fulfill the aforementioned conditions have been proposed which include: Fourier sets, finite impulse response functions, and gamma-based functions \cite{Glover_1999_H2} to mention a few. Among the alternatives, the single- and double-gamma HRFs are the most common in the literature and also integrated in the standard toolboxes \cite{Cignetti_2016_H0}. The two-gamma HRF, also known as the canonical hemodynamic response function, described in \cite{Glover_1999_H2}, models the increasing and decreasing (or undershoot) phases of the metabolic process with two gamma functions: 
\begin{align}
h\left(t\right) & =g\left(t;a_{i},\omega_{i},t_{i}\right)-g\left(t;a_{d},\omega_{d},\tau_{d}\right)\\
g\left(t;a,\omega,\tau\right) & =a\left(\frac{t}{\tau}\right)^{\kappa}\exp\left(-\kappa\frac{t-\tau}{t}\right)\label{eq:def-hrf}\\
\kappa & =\left(8\ln2\right)\left(\frac{\tau}{\omega}\right)^{2}
\end{align}
where $g\left(\cdot\right)$ is the gamma function characterized by three parameters: response height $a$, time-to-peak $\tau$, and dispersion or full-width at half-maximum $\omega$ as defined in \cite{penny2011statistical_H5,Yan_2018_H12}. $g\left(a_{i},\omega_{i},t_{i}\right)$ and $g\left(a_{d},\omega_{d},t_{d}\right)$ model the increasing and decreasing, or undershoot, phases of a typical hemodynamics.

A visual description of the six parameters (three per gamma function) of the canonical HRF with the standard values for sensorimotor activities ($a_{i}=1$, $\omega_{i}=5.2$, $\tau_{i}=5.4$, $a_{d}=0.35$, $\omega_{d}=10.8$, $\tau_{d}=7.35$ \cite{Yan_2018_H12}) is shown in Figure \ref{fig:Simulation-GLM}a. It should be denoted that these reference values are not universally accepted; some authors argue that the canonical HRF should be a double-gamma with its peak delay at, approximately, six seconds, undershoot delay of sixteen seconds, and with a peak-undershoot amplitude ratio of six \cite{penny2011statistical_H5}. However, in this paper we rely on those parameter values as in \cite{Glover_1999_H2} and \cite{Yan_2018_H12}.

From the abovementioned model, two fundamental properties should be emphasized: a) the convolution $u\left(t\right)\ast h\left(t\right)$ produces, under certain conditions, a curve that is approximately another gamma-function with a dispersion proportional to the stimulus duration \cite{penny2011statistical_H5,Blahak_2010_G1}, and b) the canonical HRF can be considered as an extension of a single gamma function model with a negative single gamma function explaining the undershoot phase. When there is no apparent undershoot in the observations, the single gamma model can also provide a good fit \cite{Boynton_1996_H10,Yan_2018_H12}. Thus, it is reasonable to conceive that the combination of GLM and HRF (GLM+HRF) can be reformulated as a concatenation of individual successive gamma functions with duration proportional to the span of each impulse, and with positive or negative amplitudes.

Relying on the same principles, we propose a hemodynamic decomposition model (HDM) where the observed hemodynamic signals (sampled every $T_{h}$ seconds) are represented as a sequence of a finite and unknown $K$ number of triggered responses $h_{k}\left(t\right)$:
\begin{align}
y\left(t\right) & =\sum_{k=1}^{K}h_{k}\left(t\right)+\varepsilon\left(t\right)\label{eq:model}\\
h_{k}\left(t\right) & =a_{k}g_{0}\left(\frac{t-\tau_{k}+\tau_{0}}{\nicefrac{\omega_{k}}{\omega_{0}}}\right)
\end{align}
where $y\left(t\right)$ is the autocorrelation effect, $h_{k}\left(t\right)$ is the triggered responses, $\varepsilon\left(t\right)$ is the random fluctuations, and $g_{0}\left(t\right)$ is a the positive gamma cycle of a sensorimotor standard HRF, for instance, $g_{0}\left(t\right)=g\left(t;a_{0}=1,\omega_{0}=5.2,\tau_{0}=5.4\right)$. Furthermore, the random fluctuations (noise) are Gaussian-distributed ($\epsilon\left(t\right)\sim\mathcal{N}\left(0,\sigma_{\varepsilon}^{2}\right)$): $\varepsilon\left(t\right)=\theta_{\varepsilon}y\left(t-T_{h}\right)+\epsilon\left(t\right)$.

HDM models the measured signal as a sum of shifted, expanded, and scaled kernel waves. Semantically, this representation is not in conflict with  the foundational claims of the GLM+HRF model. But, HDM enables representing those circumstances when the stimuli do not generate any noticeable reaction or when the response appears before or after the stimulus.

In order to describe the model in a formal framework, let us define a set $\Omega=\left\{ \left(a_{k},\omega_{k},\tau_{k}\right)\right\} _{k=1}^{K}$ that represents the three main parameters of an HRF: amplitude, duration, and starting time, respectively. The conditional log-likelihood of $N+1$ sampled points from $y\left(NT_{h}\right)$, $y\left(T_{h}\right),\ldots,y\left(\left(N-1\right)T_{h}\right)$ $\theta_{\varepsilon}$, (Equation \eqref{eq:model}) and with respect to a known initial point $y_{0}$ and $\sigma_{\varepsilon}^{2}$ is given as follows
\begin{align}
& \log\mathcal{L}\left(\Omega,\theta_{\varepsilon},\sigma_{\varepsilon}^{2}\right) = \nonumber\\
& =\log f_{\Omega,\theta_{\varepsilon},\sigma_{\varepsilon}^{2}, y\left(0\right)=y_{0}}\left(y\left(\left(N-1\right)T_{h}\right),\ldots,y\left(0\cdot T_{h}\right)\right)\nonumber \\
 & =-\frac{N}{2}\log\left(2\pi\right)-\frac{N}{2}\log\sigma_{\varepsilon}^{2}-\frac{1}{2\sigma_{\varepsilon}^{2}}\sum_{n=1}^{N}\left(S_n^*\right)^{2},\label{eq:log-likelihood}\\
&S_n^*=\delta_n-\theta_{\varepsilon}y\left(\left(n-1\right)T_{h}\right) \label{eq:log-likelihood-S}\\
&\delta_n=y\left(nT_{h}\right)-\sum_{k=1}^{K}h_{k}\left(nT_{h}\right) \label{eq:log-likelihood-deltaS}
\end{align}
The conditional maximum likelihood estimator $\left(\Omega,\theta_\varepsilon,\sigma^2_\varepsilon\right)$ can be obtained by minimizing $\log\mathcal{L}\left(\Omega,\theta_{\varepsilon},\sigma_{\varepsilon}^{2}\right)$. Details about the numerical optimization for this estimator is given in the next subsection.

\begin{center}
\begin{figure*}[th]
\begin{centering}
\includegraphics[width=0.95\textwidth]{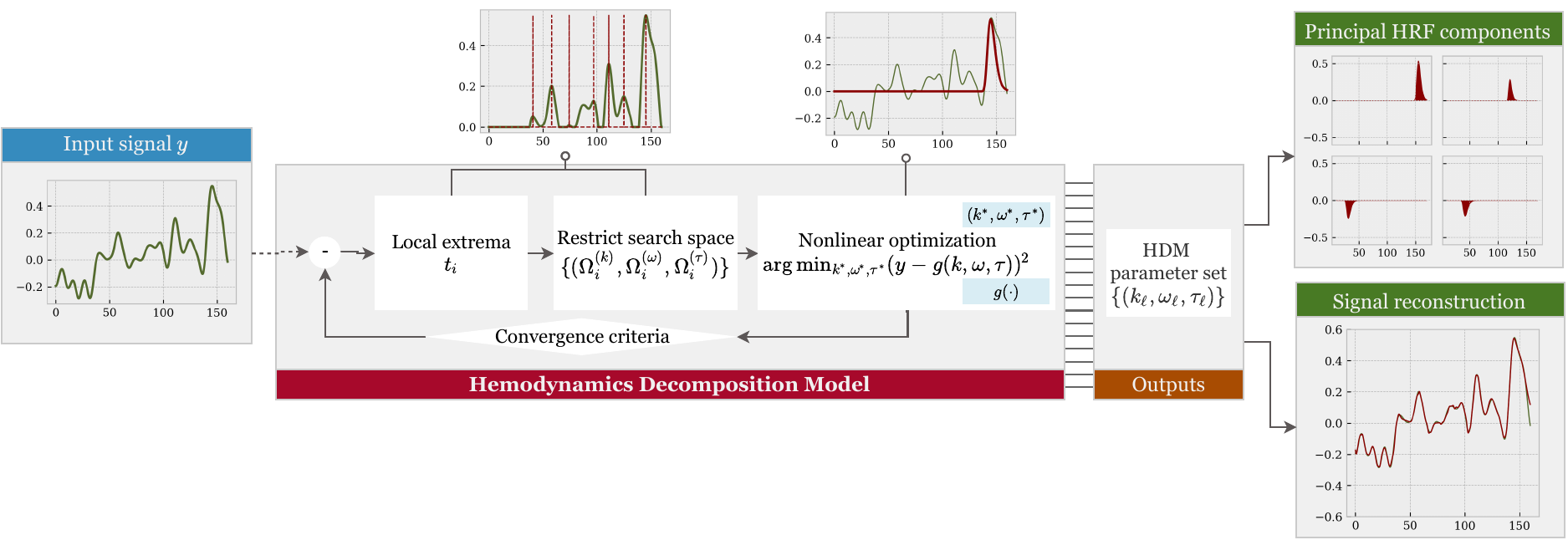}
\par\end{centering}
\caption{\label{fig:HDM-Process}Hemodynamic chartflow decomposition process. The input signal is observed (left-hand panel), and the HDM initialized (decomposing the local extrema and restrict search space, centered-top panels) then a nonlinear optimization is conduced seeking the convergence criteria, resulting in signal partitions (right-hand panels).}
\end{figure*}
\par\end{center}

\subsection{Numerical optimization algorithm}
By analogy to a mixture model with an $L_{2}$ distance, the MLE estimators of the model described in Equation \ref{eq:log-likelihood} can have several local optima, and therefore, gradient descent methods are not the most effective numerical techniques to estimate the parameters. First, we expect that a low pass filter will already remove the high-frequency noise during the preprocessing of fNIRS. Therefore, we focus on modeling the low-frequency random fluctuations ($\theta_{\varepsilon}>0$.) Hence,
\begin{align}
\hat{\Omega} &= \arg\min_\Omega \sum_{n=1}^N\left(S_n^*\right)^{2} 
=  \arg\min_\Omega\sum_{n=1}^N \left(\delta_n\right)^{2}
\end{align}
where $S_n^*$ and $\delta_n$ are defined as in Equation \ref{eq:log-likelihood-S} and Equation \ref{eq:log-likelihood-deltaS}.

Second, due to the high computational complexity, we consider $\hat{\Omega}$ to be  an accurate approximation whenever the squared Euclidean distance (or squared $L_2$ norm), between the modeled HRFs and the observed discrete signal is less than a radius of convergence $\xi$:
\begin{equation}
\sum_{n=1}^{N}\left(\delta_n\right)^{2}\le\xi
\end{equation}
Third, to ensure a consistent parameter estimation, we introduce a iterative estimation approach (Figure \ref{fig:HDM-Process}) for constructing the parameter search space:
\begin{itemize}
\item Construct $\left\{ \left(t_{i},y_{i}\right)\right\} $ as a set of $M$ local extreme points, i.e., those points such where $\left.\frac{dy}{dt}\right|_{t=t_{i}}=0$.
\item For each time index $t_{i}$, define three sets: $\Omega_{a}=\left[\kappa_{a0}y_{i},y_{i}\right]$, $\Omega_{\omega}=\left[\omega_{\epsilon},\omega_{m}\right]$, $\Omega_{\tau}=\left[\kappa_{\tau0}\tau_{0},\kappa_{\tau1}\tau_{0}\right]$. $\Omega_{\tau}$ describes the potential ranges where the hemodynamic impulse are expected to be produced. According to \cite{penny2011statistical_H5} and \cite{Yan_2018_H12}, the common starting time is located from five to sixteen seconds before an observed peak ($\kappa_{\tau0}=0.8$, $\kappa_{\tau0}=3$). $\Omega_{\omega}$ outlines the feasible choices for the duration. Based on \cite{penny2011statistical_H5} and \cite{Glover_1999_H2}, the maximum observed duration experimentally is slightly higher than seven seconds in sensorimotor stimuli; therefore, we set $\omega_{m}=8$. The lowerbound $\omega_{\epsilon}$ was empirically set to 0.1 seconds in our experiments. Finally, $\Omega_{a}$ denotes the search space of HRF amplitudes. We prioritize those components that explain most of the observed amplitudes. Consequently, we define the upperbound as the observed amplitude $y_{i}$, and the lower limit as a proportion of $y_{i}$. Furthermore, in our analysis we set $\kappa_{a0}=0.8$.
\item For each triplet $\left(\Omega_{a},\Omega_{\omega},\Omega_{\tau}\right)$ related to the time $t_{i}$, we use the Broyden--Fletcher--Goldfarb--Shanno optimization algorithm to optimize the best parameters $\left(a^{*},\omega^{*},\tau^{*}\right)$ that minimize the $L_{2}$ distance between the estimation and the observations in the time interval $\left[t_{i},t_{i}+2\max\left(\Omega_{\tau}\right)\right]$. Due to the waveform of the gamma-function, we submit that the component effect outside the interval will not be significant. The previous constraints not only guarantee finding the best component that locally explains the input signal, but  also reduce the computational time due to a supplementary contraction in the search space.
\item Define $R^{\left(0\right)}=y\left(t\right)$, and let $R^{\left(\ell\right)}$ and $r^{\left(\ell\right)}$ be residuals of the $\ell$-th iteration: $R^{\left(\ell\right)}=R^{\left(\ell-1\right)}-a^{*}g_{0}\left(\frac{t-a^{*}+\tau_{0}}{\nicefrac{a^{*}}{\omega_{0}}}\right)$ and $r^{\left(\ell\right)}$ the residual sum of squares $r^{\left(\ell\right)}=\left|R^{\left(\ell\right)}\right|^{2}$. The previous steps can be repeated until $r^{\left(\ell\right)}$ reaches the threshold error $\xi$ or $L$ iterations.
\item Given the set of estimators $\left\{ \left(a_{i}^{*},\omega_{i}^{*},\tau_{i}^{*}\right)\right\} _{i}$, the autocorrelation parameter $\hat{\theta}_{\varepsilon}$ is obtained using least square estimation:
\noindent\begin{align}
\hat{\theta}_{\varepsilon} & =\arg\min_{\theta_{\varepsilon}}\sum_{n=1}^{N}\left(\delta_n^*-\theta_{\varepsilon}y\left(\left(n-1\right)T_{h}\right)\right)^{2}\\
\delta_n^*&=y\left(nT_{h}\right)-\sum_{k=1}^{K}a_{k}^{*}g_{0}\left(\frac{nT_{h}-\tau_{k}^{*}+\tau_{0}}{\nicefrac{\omega_{k}^{*}}{\omega_{0}}}\right)
\end{align}

\end{itemize}

\section{Experimental analysis}

\begin{figure}[t]
\begin{centering}
\includegraphics[width=0.95\columnwidth,height=5cm,keepaspectratio]{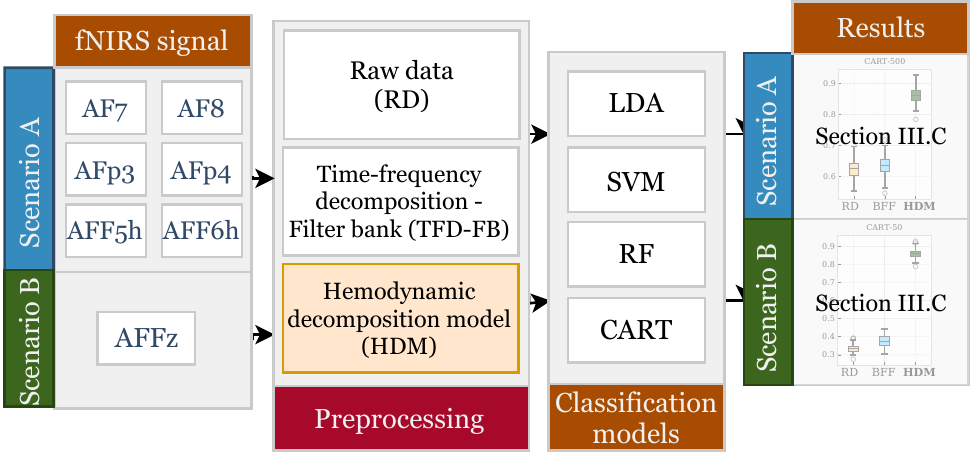}
\par\end{centering}
\caption{\label{fig:chart} Visual summary of the experimental framework.}
\end{figure}

In this section, we explore the potential of our proposed decomposition technique --HDM--, in the role of the methodological framework combined with the feature extraction method in a machine learning context as it is illustrated in Figure \ref{fig:chart}. We define a classification problem with the goal of predicting the cognitive level related to an activity that a subject is performing based on his hemodynamic signals. 

\subsection{Dataset}
\begin{figure}[t]
\begin{centering}
\includegraphics[width=0.95\textwidth]{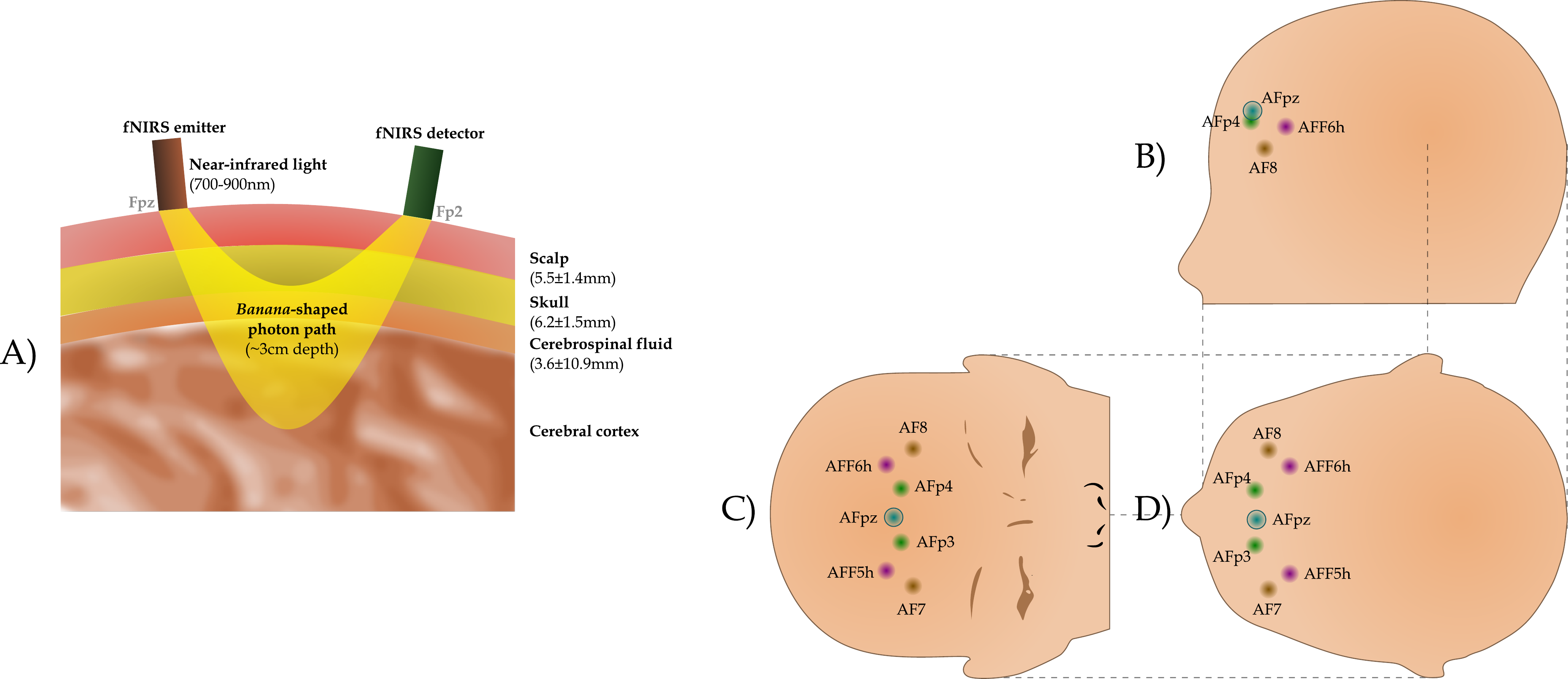}
\par\end{centering}
\caption{\label{fig:def-and-locations}Principles and measurement settings of fNIRS. (A) Illustration of the the working principle of fNRIS, that is, a Physiological channels used in the analysis. The selected seven fNIRS channels, placed in the frontal cortex, are showed in panels (B) side, (C) front, and (D) top view.}
\end{figure}

The dataset of the experiment comprises oxy- and deoxy-hemoglobin signals estimated from the fNIRS using the modified Beer-Lambert law. For repeatability purposes, a publicly available dataset is being used for our pilot analysis which comes from a series of experiments performed by Shin et al. at the Technische Universit{\"a}t Berlin involving \textquotedbl n-back\textquotedbl{} tasks \cite{Shin_2018_ML2}. Their study was restricted to 29 subjects in the age group of 20-30 years old. We rather focused on one of their experiments: n-back tasks, which are accurate markers of brain workload \cite{Whiteman_2017_ML1}.
The dataset contains a series of sessions, where each session protocol is composed of four steps: a) show a straightforward instruction for 2 seconds; b) request the subject to perform the task (0-, 2-, and 3-back); c) ring a short beep for 250ms; and d) rest for 20 seconds. For further information about the experimental protocol, we refer the reader to \cite{Shin_2018_ML2}. The fNIRS signals in the dataset were sampled at 10Hz on 36 channels. For the purpose of this study, we selected seven channels in the frontal cortex: AF7, AF8, AFF5h, AFF6h, AFp3, AFp4, and AFpz, according to the EEG 10-20 standard layout position, as it is shown in Figure \ref{fig:def-and-locations} \cite{Shin_2018_ML2}.

\subsection{Experimental setting}

Based on the complete dataset provided by Shin et al. \cite{Shin_2018_ML2}, we configured two simulation scenarios to test our method:
\begin{itemize}
\item Scenario A: classification using six channels from the frontal region: AF7, AF8, AFF5, AFF6, AFp3, AFp4 (Figure \ref{fig:def-and-locations}) with data from the \textquotedblleft subject 01.\textquotedblright{} This setting will expose the ability of the hemodynamic decomposition method to provide relevant information in machine learning algorithms as it is known that cognitive load manifests with activations in the frontal cortex \cite{racz2017increased}.
\item Scenario B: workload identification using only the channel AFpz (frontal region slightly over the root of the nose, Figure \ref{fig:def-and-locations}). In contrast with the previous simulation, due to the reduced amount of data used in this setting, this scenario is assessed with data from subjects 01-26. The results of this simulation will assess the capabilities of our method under spatially limited data.
\end{itemize}
To compare the efficiency of our model as a pre-processing technique, we establish two alternative feature sets: raw time-domain signals, and time-frequency features. The time-frequency decomposition (TFD) is based on a filter bank structure (FB): we define five 100th-order finite-impulse-response band-pass filters in the ranges: 0-0.02Hz, 0.02-0.04Hz, 0.06-0.08Hz, and 0.08-0.10Hz. These intervals are known to contain the most relevant information related to hemodynamic processes, and Appriou et al. showed that the TFD-FB provides a prediction accuracy of 67.9\% $\pm$ 8.1\% \cite{Appriou_2018_ML7}.

\begin{figure*}[th]
\begin{centering}
\includegraphics[width=0.95\textwidth]{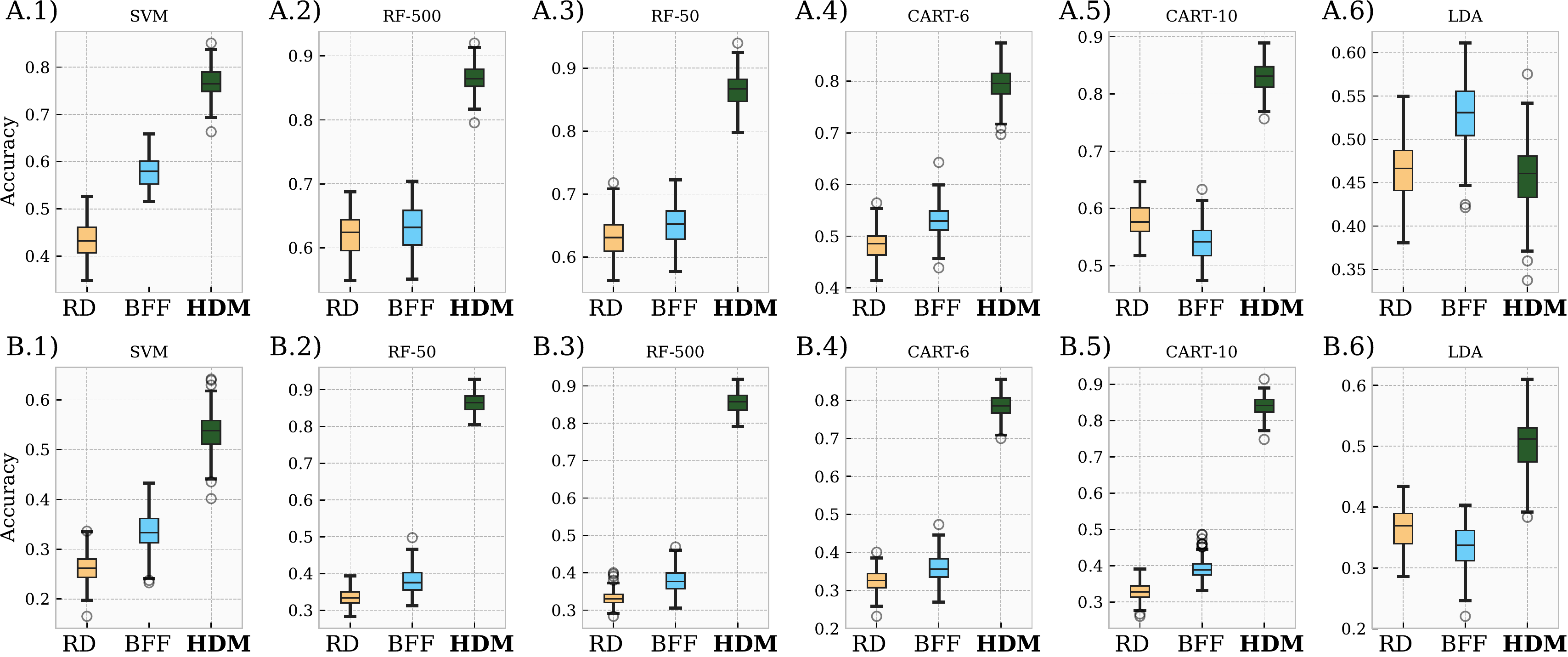}
\par\end{centering}
\caption{\label{fig:boxplot-results}Performance of HDM in activity recognition on the two different brain region, adopting six machine-learning algorithms, based on three data pre-processing. Scenario A (top panels): data gathered from the six frontal-cortex channels: AF7, AF8, AFF5, AFF6, AFp3, and AFp4 of subject 01. Scenario B (bottom panels): data collected from subject 01 from only channel AFpz.\protect \\
Six machine-learning algorithms were compared (each panel from the left to the right): 1) SVM; 2) random forest with 50 trees and 3) 500 trees; 4) decision trees (CART) with 6 depth levels and 5) 10 levels; and 6) LDA classifier.\protect \\
The abscissa axis represents the type of input on the algorithms: raw data (RD), bank filter features (BFF), and HDM. In all evaluations, box plots show the second and third quantile of the results using SWTT. Empirical results showed the outperformance of the HDM across almost all classifiers.}
\end{figure*}

For each feature extraction technique, we applied four basic machine algorithms and compare their performance: support vector machines (SVM) \cite{Fernandez_Rojas_2019_A,Putze_2014_C}, random forests (RF) \cite{Erdo_an_2019_ML3,Keshmiri_2017_ML4} classification and regression trees (CART) \cite{Sirpal_2019_ML5}, and linear discriminant analysis (LDA) \cite{Appriou_2018_ML7}. These algorithms were selected, given their proven efficiency to solve other classification tasks were fNIRS was involved. Based on the literature results \cite{Appriou_2018_ML7}, there were no significant differences between the performance of recursive neural networks respect with LDA with features extracted with filter banks. Due to this reason and the lack of longer time series per subject, deep learning routines were not included in this comparison. For further details about the applied algorithms, we refer to \cite{Fernandez_Rojas_2019_A,Keshmiri_2017_ML4,Sirpal_2019_ML5,Appriou_2018_ML7}.

Physiological signals hold substantial information about their time dependency structure. In order to keep this structure in our experimental setting, the validation method consisted of a sliding window train-test (SWTT) \cite{bouktif2019single}: the algorithm predicted the following activities in 2 seconds using the information from the previous 60 seconds. This estimation technique is known for being unbiased with a moderately low variance \cite{kohavi1995study_EST1}.

\subsection{Results}
\begin{figure*}[th]
\begin{center}
\noindent\includegraphics[width=0.95\textwidth]{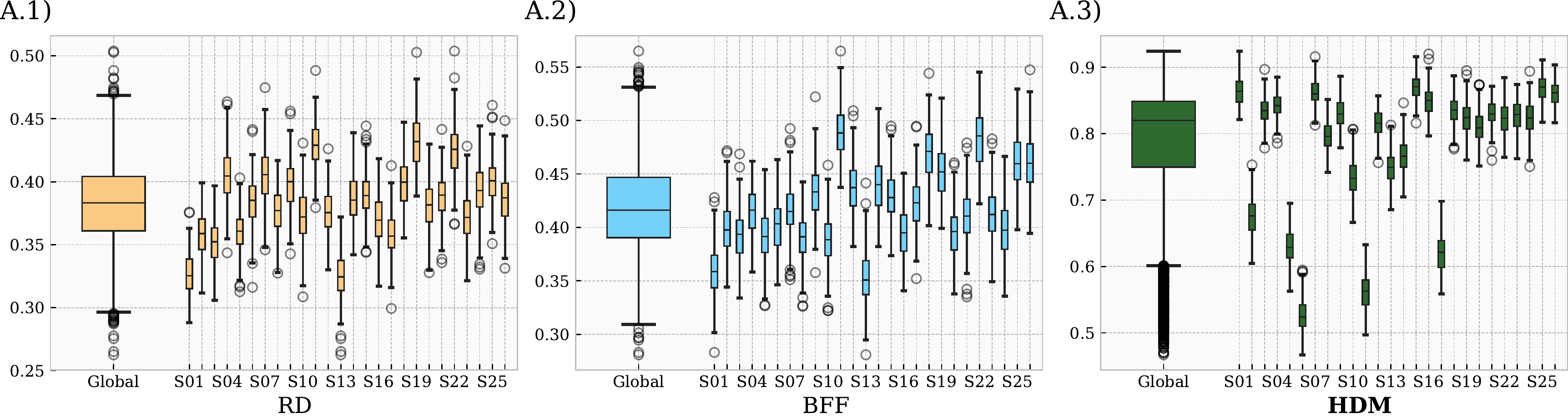}\\%
\noindent\includegraphics[width=0.95\textwidth]{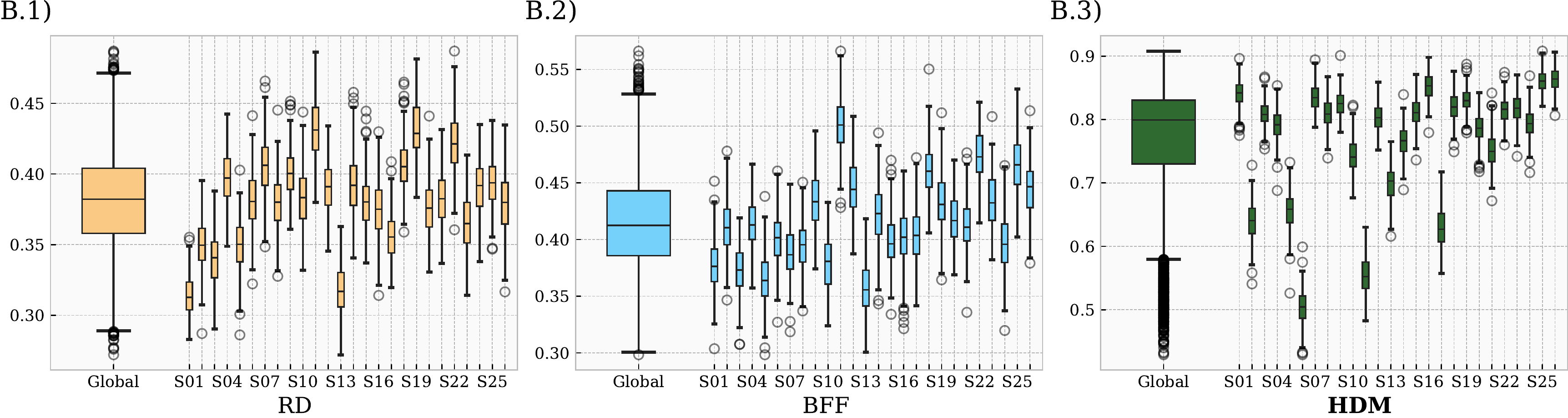}
\caption{\label{fig:cross-subjects-boxplot-results}Subject-level performance of HDM in a single-channel activity-recognition in 26 subjects. Three preprocessing methods were compared: 1) raw data (RD), left-hand panels, and bank filter features (BFF), center panels, and 3) HDM, right-hand panels. Two algorithms were applied: A) random forest with 50 trees (top panels); B) Decision trees CART --10-level depth-- (bottom panels). The identifier of the subject from whom data is collected is shown in the x-axis for each plot. Note that HDM provides a higher accuracy with both classification models. }
\end{center}
\end{figure*}

Since the distribution of the four labels  (0-back, 2-back, 3-back, and rest) is balanced across the dataset, the accuracy is a proper metric of the algorithms' performance, and thus it is used to compare the outcomes. We compiled the results of the experimental scenario A (with frontal cortex data) and B (with single-channel signals) from ``subject 01'' in Figure \ref{fig:boxplot-results}.

From the empirical results, there are a few noteworthy observations regarding the performance of the different machine learning algorithms:
\begin{itemize}
\item Random forests showed more success at identifying cognitive load states regardless of the experimental setting. These results are also concordant with the conclusions found in previous studies \cite{Appriou_2018_ML7}.
\item Support vector machines (SVM) did not exhibit the best performance in our experiments. We examine this behavior using the cumulative confusion matrices of Table \ref{tab:confusion-matrix-A} and \ref{tab:confusion-matrix-B}. Even though SVM performs worse than the other algorithms, it can recognize the n-back activities, but it exhibits a high level of uncertainty when differentiating those states from their respective rest periods. However, the level of uncertainty among n-back activities is minimal when using HDM with respect to TFD-FB or raw data.
\item By comparing experiments A and B, it is evident that the performance drops when the input data is restricted to one channel. The only two exceptions are linear discriminant analysis and random forests with the decomposition model HDM. The sensibility of LDA improves by 8.88\% and its precision by 11.1\%. Random forests maintain identical performance across both cases.
\item Decision trees do not exhibit a substantial difference with the SVM alternatives where TFD-FB and raw data are the input features. Nevertheless, they yield attractive accuracy metrics when HDM is utilized: between 73.2\% and 83.8\%. HDM produces a non-linearly mapping of each task into hemodynamic responses, and we hypothesize that decision trees are partially detecting those non-linear boundaries. 
\end{itemize}
The results indicate that the increment of the prediction power using HDM is not intrinsically associated with one specific family of ML algorithms. We can argue that {HDM is a proper method to extract relevant features from the fNIRS signals. However, the precision of brain activity detection will rely on the chosen algorithm}. Additionally, the remarkable results of the decision trees in experiment B are promissory for further development of fNIRS activity recognition systems, especially implemented in devices with hardware restrictions or real-time architectures (due to the low time complexity of training CART methods).

\section{Conclusion}

We propose an appealing representation of fNIRS signals using a hemodynamic decomposition model (HDM) that has an intuitive physiological interpretation. It models the observed signal as a sum of independent hemodynamic responses, where each one is based on the canonical HRF, but rescaled in duration and amplitude, and triggered at different instants. Estimating those parameters and the number of HRFs that accurately model the hemodynamic signals is time-exhaustive, nonconvex optimization problem. To cope with this problem, we present an iterative estimation process that maximizes the accuracy of the signal and allows us to control the precision of the representation using only two hyper-parameters: the number of maximum iterations and the convergence ratio.

Due to the data-driven nature of the HDM model, no additional information about the stimuli in the experiment is required. Nevertheless, the method can be combined with registered stimuli as a feature extraction method for machine learning algorithms in prediction/classification tasks. We investigated the case of brain cognition load prediction under a classical experiment setting (n-back). Raw data and time-frequency features were compared with our decomposition model using different classification methods including support vector machines, decision treees and random forests. We conclude that our method has a clear advantage as a feature transformation regardless of the deployed algorithm. HDM yields an improvement in the accuracy by up to 21.59\% and 23.69\% with respect to time-spectral and raw data features, respectively.

\section{Funding}
This work is financially supported by the Research Council of Norway to the Project No. 273599, "Patient-Centric Engineering in Rehabilitation (PACER)". Diego C. Nascimento is partially supported by the Brazilian funding agency "Coordena\c{c}\~{a}o de Aperfei\c{c}oamento de Pessoal de N\'{i}vel Superior" (CAPES).

\bibliographystyle{IEEEtran}
\bibliography{references}

\onecolumn\clearpage{}

\appendix

\section{Performance results}

\exhyphenpenalty=10000\relax{}

\begin{table}[H]
\caption{\label{tab:performances}Performance metrics for scenarios A and B: sensibility (SENS), specificity (SPEC), and general accuracy (ACC).}

\centering\scalebox{0.9}{

\footnotesize%
\begin{tabular}{|>{\raggedright}m{1.3cm}|>{\raggedright}m{1cm}|>{\centering}m{0.7cm}>{\centering}m{0.7cm}|>{\centering}m{0.7cm}>{\centering}m{0.7cm}|>{\centering}m{0.7cm}>{\centering}m{0.7cm}|>{\centering}m{0.7cm}>{\centering}m{0.7cm}|>{\centering}m{0.7cm}>{\centering}m{0.7cm}|>{\centering}m{0.7cm}>{\centering}m{0.7cm}|}
\hline 
 &  & \multicolumn{6}{c|}{Scenario A} & \multicolumn{6}{c|}{Scenario B}\tabularnewline
\cline{3-14} 
Classifier & Activity & \multicolumn{2}{c|}{\textbf{Raw data}} & \multicolumn{2}{c|}{\textbf{TFD-FB}} & \multicolumn{2}{c|}{\textbf{HDM}} & \multicolumn{2}{c|}{\textbf{Raw data}} & \multicolumn{2}{c|}{\textbf{TFD-FB}} & \multicolumn{2}{c|}{\textbf{HDM}}\tabularnewline
\cline{3-14} 
 &  & SENS & PREC & SENS & PREC & SENS & PREC & SENS & PREC & SENS & PREC & SENS & PREC\tabularnewline
\hline 
 & Rest & 0.538 & 0.589 & 0.611 & 0.664 & 0.669 & 0.627 & 0.469 & 0.401 & 0.469 & 0.336 & 0.466 & 0.518\tabularnewline
\textbf{SVM-L} & 0-back & 0.521 & 0.44 & 0.598 & 0.592 & 0.606 & 0.684 & 0.151 & 0.156 & 0.151 & 0.304 & 0.469 & 0.362\tabularnewline
 & 2-back & 0.399 & 0.345 & 0.628 & 0.478 & 0.787 & 0.734 & 0.182 & 0.199 & 0.182 & 0.134 & 0.483 & 0.38\tabularnewline
 & 3-back & 0.456 & 0.5 & 0.517 & 0.55 & 0.731 & 0.815 & 0.178 & 0.228 & 0.178 & 0.142 & 0.429 & 0.535\tabularnewline
 & \textbf{ACC} & \textbf{0.498} &  & \textbf{0.596} &  & \textbf{0.687} &  & \textbf{0.287} &  & \textbf{0.287} &  & \textbf{0.462} & \tabularnewline
\hline 
 & Rest & 0.500 & 0.528 & 0.604 & 0.674 & 0.739 & 0.74 & 0.417 & 0.317 & 0.417 & 0.389 & 0.531 & 0.608\tabularnewline
\textbf{SVM-R} & 0-back & 0.483 & 0.47 & 0.617 & 0.599 & 0.656 & 0.715 & 0.227 & 0.302 & 0.227 & 0.515 & 0.491 & 0.513\tabularnewline
 & 2-back & 0.298 & 0.253 & 0.630 & 0.455 & 0.893 & 0.774 & 0.131 & 0.166 & 0.131 & 0.26 & 0.695 & 0.465\tabularnewline
 & 3-back & 0.347 & 0.371 & 0.476 & 0.5 & 0.818 & 0.865 & 0.154 & 0.144 & 0.154 & 0.06 & 0.512 & 0.481\tabularnewline
 & \textbf{ACC} & \textbf{0.439} &  & \textbf{0.590} &  & \textbf{0.760} &  & \textbf{0.258} &  & \textbf{0.258} &  & \textbf{0.542} & \tabularnewline
\hline 
 & Rest & 0.632 & 0.724 & 0.657 & 0.73 & 0.855 & 0.848 & 0.444 & 0.454 & 0.444 & 0.527 & 0.855 & 0.848\tabularnewline
\textbf{RF-50} & 0-back & 0.652 & 0.568 & 0.593 & 0.567 & 0.858 & 0.874 & 0.234 & 0.229 & 0.234 & 0.223 & 0.857 & 0.873\tabularnewline
 & 2-back & 0.590 & 0.473 & 0.627 & 0.58 & 0.877 & 0.877 & 0.247 & 0.256 & 0.247 & 0.339 & 0.877 & 0.877\tabularnewline
 & 3-back & 0.654 & 0.647 & 0.706 & 0.576 & 0.889 & 0.889 & 0.226 & 0.206 & 0.226 & 0.175 & 0.889 & 0.889\tabularnewline
 & \textbf{ACC} & \textbf{0.632} &  & \textbf{0.646} &  & \textbf{0.865} &  & \textbf{0.334} &  & \textbf{0.334} &  & \textbf{0.865} & \tabularnewline
\hline 
 & Rest & 0.638 & 0.723 & 0.670 & 0.73 & 0.855 & 0.848 & 0.445 & 0.448 & 0.445 & 0.52 & 0.855 & 0.848\tabularnewline
\textbf{RF-100} & 0-back & 0.649 & 0.585 & 0.638 & 0.575 & 0.858 & 0.874 & 0.234 & 0.232 & 0.234 & 0.231 & 0.858 & 0.874\tabularnewline
 & 2-back & 0.594 & 0.488 & 0.634 & 0.616 & 0.877 & 0.877 & 0.246 & 0.255 & 0.246 & 0.341 & 0.877 & 0.877\tabularnewline
 & 3-back & 0.651 & 0.626 & 0.717 & 0.643 & 0.889 & 0.889 & 0.221 & 0.209 & 0.221 & 0.18 & 0.889 & 0.889\tabularnewline
 & \textbf{ACC} & \textbf{0.635} &  & \textbf{0.664} &  & \textbf{0.865} &  & \textbf{0.332} &  & \textbf{0.332} &  & \textbf{0.865} & \tabularnewline
\hline 
 & Rest & 0.587 & 0.572 & 0.620 & 0.619 & 0.763 & 0.815 & 0.417 & 0.479 & 0.417 & 0.466 & 0.749 & 0.749\tabularnewline
\textbf{CART-6} & 0-back & 0.443 & 0.341 & 0.491 & 0.472 & 0.808 & 0.745 & 0.207 & 0.179 & 0.207 & 0.259 & 0.837 & 0.678\tabularnewline
 & 2-back & 0.365 & 0.401 & 0.471 & 0.524 & 0.850 & 0.764 & 0.219 & 0.212 & 0.219 & 0.271 & 0.598 & 0.69\tabularnewline
 & 3-back & 0.430 & 0.537 & 0.467 & 0.42 & 0.807 & 0.829 & 0.261 & 0.199 & 0.261 & 0.271 & 0.770 & 0.805\tabularnewline
 & \textbf{ACC} & \textbf{0.488} &  & \textbf{0.542} &  & \textbf{0.793} &  & \textbf{0.325} &  & \textbf{0.325} &  & \textbf{0.732} & \tabularnewline
\hline 
\multirow{1}{1.3cm}{} & Rest & 0.663 & 0.679 & 0.646 & 0.598 & 0.821 & 0.842 & 0.427 & 0.457 & 0.427 & 0.521 & 0.836 & 0.76\tabularnewline
\textbf{CART-10} & 0-back & 0.550 & 0.468 & 0.509 & 0.529 & 0.828 & 0.84 & 0.213 & 0.208 & 0.213 & 0.271 & 0.845 & 0.786\tabularnewline
 & 2-back & 0.541 & 0.51 & 0.474 & 0.596 & 0.889 & 0.832 & 0.225 & 0.214 & 0.225 & 0.373 & 0.683 & 0.869\tabularnewline
 & 3-back & 0.482 & 0.577 & 0.428 & 0.358 & 0.839 & 0.829 & 0.235 & 0.206 & 0.235 & 0.198 & 0.885 & 0.889\tabularnewline
 & \textbf{ACC} & \textbf{0.589} &  & \textbf{0.548} &  & \textbf{0.838} &  & \textbf{0.323} &  & \textbf{0.323} &  & \textbf{0.806} & \tabularnewline
\hline 
 & Rest & 0.583 & 0.574 & 0.709 & 0.556 & 0.629 & 0.381 & 0.410 & 0.652 & 0.410 & 0.534 & 0.613 & 0.401\tabularnewline
\textbf{LDA} & 0-back & 0.540 & 0.477 & 0.609 & 0.564 & 0.416 & 0.568 & 0.215 & 0.131 & 0.215 & 0.207 & 0.566 & 0.657\tabularnewline
 & 2-back & 0.327 & 0.316 & 0.386 & 0.477 & 0.380 & 0.455 & 0.249 & 0.133 & 0.249 & 0.162 & 0.554 & 0.533\tabularnewline
 & 3-back & 0.347 & 0.433 & 0.375 & 0.542 & 0.375 & 0.548 & 0.257 & 0.084 & 0.257 & 0.129 & 0.422 & 0.804\tabularnewline
 & \textbf{ACC} & \textbf{0.483} &  & \textbf{0.540} &  & \textbf{0.457} &  & \textbf{0.362} &  & \textbf{0.362} &  & \textbf{0.537} & \tabularnewline
\hline 
\end{tabular}\normalfont

}
\end{table}

\clearpage{}

\section{Confusion matrices for simulations}

\begin{table}[H]
\caption{\label{tab:confusion-matrix-A}Confusion matrix for simulated scenario A (six frontal-cortex channels)}

\centering\scalebox{0.9}{

\footnotesize%
\begin{tabular}{|>{\raggedright}m{1.3cm}|>{\raggedright}m{1cm}|>{\centering}m{0.5cm}>{\centering}m{0.7cm}>{\centering}m{0.7cm}>{\centering}m{0.7cm}|>{\centering}m{0.65cm}>{\centering}m{0.7cm}>{\centering}m{0.7cm}>{\centering}m{0.7cm}|>{\centering}m{0.65cm}>{\centering}m{0.75cm}>{\centering}m{0.75cm}>{\centering}m{0.75cm}|}
\hline 
 &  & \multicolumn{4}{c|}{\textbf{Raw data}} & \multicolumn{4}{c|}{\textbf{BFF}} & \multicolumn{4}{c|}{\textbf{HDM}}\tabularnewline
\cline{3-14} 
Classifier & Actual & \multicolumn{4}{c|}{Predicted activity} & \multicolumn{4}{c|}{Predicted activity} & \multicolumn{4}{c|}{Predicted activity}\tabularnewline
\cline{3-14} 
 & activity & Rest & 0-back & 2-back & 3-back & Rest & 0-back & 2-back & 3-back & Rest & 0-back & 2-back & 3-back\tabularnewline
\hline 
 & Rest & 4336 & 976 & 1184 & 863 & 4889 & 868 & 698 & 904 & 4612 & 1395 & 636 & 716\tabularnewline
\textbf{SVM-L} & 0-back & 1533 & 1381 & 159 & 68 & 1120 & 1859 & 123 & 39 & 991 & 2150 & 0 & 0\tabularnewline
 & 2-back & 1311 & 285 & 1105 & 499 & 1062 & 317 & 1531 & 290 & 847 & 0 & 2348 & 5\tabularnewline
 & 3-back & 873 & 8 & 319 & 1200 & 933 & 63 & 84 & 1320 & 444 & 0 & 0 & 1956\tabularnewline
\hline 
 & Rest & 3888 & 1178 & 1164 & 1129 & 4961 & 778 & 643 & 977 & 5442 & 1176 & 296 & 445\tabularnewline
\textbf{SVM-R} & 0-back & 1292 & 1476 & 309 & 64 & 1089 & 1882 & 122 & 48 & 895 & 2246 & 0 & 0\tabularnewline
 & 2-back & 1591 & 317 & 808 & 484 & 1118 & 328 & 1456 & 298 & 707 & 0 & 2476 & 17\tabularnewline
 & 3-back & 1000 & 82 & 428 & 890 & 1046 & 63 & 90 & 1201 & 323 & 0 & 0 & 2077\tabularnewline
\hline 
 & Rest & 5327 & 764 & 682 & 586 & 5375 & 819 & 686 & 479 & 6243 & 456 & 393 & 267\tabularnewline
\textbf{RF-50} & 0-back & 1069 & 1783 & 270 & 19 & 1027 & 1781 & 326 & 7 & 397 & 2744 & 0 & 0\tabularnewline
 & 2-back & 1285 & 185 & 1513 & 217 & 987 & 269 & 1855 & 89 & 393 & 0 & 2807 & 0\tabularnewline
 & 3-back & 747 & 2 & 98 & 1553 & 790 & 134 & 93 & 1383 & 267 & 0 & 0 & 2133\tabularnewline
\hline 
 & Rest & 5317 & 778 & 679 & 585 & 5370 & 733 & 698 & 558 & 6243 & 456 & 393 & 267\tabularnewline
\textbf{RF-100} & 0-back & 988 & 1838 & 291 & 24 & 982 & 1806 & 353 & 0 & 397 & 2744 & 0 & 0\tabularnewline
 & 2-back & 1245 & 200 & 1560 & 195 & 973 & 206 & 1970 & 51 & 393 & 0 & 2807 & 0\tabularnewline
 & 3-back & 785 & 14 & 98 & 1503 & 687 & 86 & 85 & 1542 & 267 & 0 & 0 & 2133\tabularnewline
\hline 
 & Rest & 4210 & 1056 & 1349 & 744 & 4556 & 1183 & 1007 & 613 & 5997 & 548 & 432 & 382\tabularnewline
\textbf{CART-6} & 0-back & 1353 & 1072 & 548 & 168 & 1008 & 1484 & 518 & 131 & 802 & 2339 & 0 & 0\tabularnewline
 & 2-back & 907 & 213 & 1283 & 797 & 955 & 159 & 1678 & 408 & 652 & 9 & 2446 & 93\tabularnewline
 & 3-back & 700 & 78 & 333 & 1289 & 834 & 195 & 363 & 1008 & 410 & 0 & 0 & 1990\tabularnewline
\hline 
\multirow{1}{1.3cm}{} & Rest & 4998 & 942 & 707 & 712 & 4398 & 1165 & 1175 & 621 & 6197 & 548 & 332 & 282\tabularnewline
\textbf{CART-10} & 0-back & 1017 & 1470 & 418 & 236 & 845 & 1663 & 495 & 138 & 496 & 2639 & 0 & 6\tabularnewline
 & 2-back & 836 & 191 & 1632 & 541 & 745 & 159 & 1907 & 389 & 444 & 0 & 2663 & 93\tabularnewline
 & 3-back & 689 & 69 & 258 & 1384 & 817 & 282 & 443 & 858 & 410 & 0 & 0 & 1990\tabularnewline
\hline 
 & Rest & 4222 & 777 & 1130 & 1230 & 4090 & 650 & 1264 & 1355 & 2804 & 1660 & 1494 & 1401\tabularnewline
\textbf{LDA} & 0-back & 1095 & 1499 & 459 & 88 & 507 & 1770 & 618 & 246 & 729 & 1785 & 513 & 114\tabularnewline
 & 2-back & 1205 & 350 & 1010 & 635 & 715 & 394 & 1526 & 565 & 745 & 319 & 1456 & 680\tabularnewline
 & 3-back & 721 & 149 & 492 & 1038 & 456 & 94 & 549 & 1301 & 182 & 532 & 370 & 1316\tabularnewline
\hline 
\end{tabular}\normalfont

}
\end{table}

\begin{table}[H]
\caption{\label{tab:confusion-matrix-B}Confusion matrix for simulated scenario B (single frontal-cortex channel)}

\centering\scalebox{0.9}{

\footnotesize%
\begin{tabular}{|>{\raggedright}m{1.3cm}|>{\raggedright}m{1cm}|>{\centering}m{0.5cm}>{\centering}m{0.7cm}>{\centering}m{0.7cm}>{\centering}m{0.7cm}|>{\centering}m{0.65cm}>{\centering}m{0.7cm}>{\centering}m{0.7cm}>{\centering}m{0.7cm}|>{\centering}m{0.65cm}>{\centering}m{0.75cm}>{\centering}m{0.75cm}>{\centering}m{0.75cm}|}
\hline 
 &  & \multicolumn{4}{c|}{\textbf{Raw data}} & \multicolumn{4}{c|}{\textbf{BFF}} & \multicolumn{4}{c|}{\textbf{HDM}}\tabularnewline
\cline{3-14} 
Classifier & Actual & \multicolumn{4}{c|}{Predicted activity} & \multicolumn{4}{c|}{Predicted activity} & \multicolumn{4}{c|}{Predicted activity}\tabularnewline
\cline{3-14} 
 & activity & Rest & 0-back & 2-back & 3-back & Rest & 0-back & 2-back & 3-back & Rest & 0-back & 2-back & 3-back\tabularnewline
\hline 
 & Rest & 2953 & 1529 & 1558 & 1319 & 2473 & 2457 & 1572 & 857 & 3809 & 1044 & 1094 & 1412\tabularnewline
\textbf{SVM-L} & 0-back & 1438 & 489 & 870 & 344 & 1249 & 954 & 874 & 64 & 1902 & 1137 & 102 & 0\tabularnewline
 & 2-back & 1146 & 549 & 638 & 867 & 1458 & 891 & 428 & 423 & 1451 & 239 & 1216 & 294\tabularnewline
 & 3-back & 759 & 662 & 431 & 548 & 732 & 794 & 533 & 341 & 1009 & 5 & 103 & 1283\tabularnewline
\hline 
 & Rest & 2335 & 1846 & 2153 & 1025 & 2864 & 2415 & 1037 & 1043 & 4474 & 1373 & 653 & 859\tabularnewline
\textbf{SVM-R} & 0-back & 1036 & 949 & 845 & 311 & 913 & 1619 & 557 & 52 & 1429 & 1612 & 0 & 100\tabularnewline
 & 2-back & 1315 & 791 & 532 & 562 & 1360 & 716 & 831 & 293 & 1369 & 200 & 1489 & 142\tabularnewline
 & 3-back & 908 & 603 & 544 & 345 & 861 & 933 & 461 & 145 & 1148 & 97 & 0 & 1155\tabularnewline
\hline 
 & Rest & 3339 & 1451 & 1566 & 1003 & 3877 & 1297 & 1333 & 852 & 6240 & 459 & 393 & 267\tabularnewline
\textbf{RF-50} & 0-back & 1570 & 720 & 568 & 283 & 1562 & 702 & 705 & 172 & 397 & 2743 & 1 & 0\tabularnewline
 & 2-back & 1470 & 502 & 819 & 409 & 1498 & 310 & 1084 & 308 & 393 & 0 & 2807 & 0\tabularnewline
 & 3-back & 1147 & 399 & 359 & 495 & 1062 & 543 & 376 & 419 & 267 & 0 & 0 & 2133\tabularnewline
\hline 
 & Rest & 3297 & 1458 & 1551 & 1053 & 3825 & 1323 & 1296 & 915 & 6243 & 456 & 393 & 267\tabularnewline
\textbf{RF-100} & 0-back & 1540 & 728 & 587 & 286 & 1558 & 727 & 697 & 159 & 397 & 2744 & 0 & 0\tabularnewline
 & 2-back & 1445 & 509 & 816 & 430 & 1466 & 329 & 1091 & 314 & 393 & 0 & 2807 & 0\tabularnewline
 & 3-back & 1128 & 413 & 358 & 501 & 1035 & 556 & 376 & 433 & 267 & 0 & 0 & 2133\tabularnewline
\hline 
 & Rest & 3522 & 1449 & 1500 & 888 & 3430 & 1491 & 1232 & 1206 & 5509 & 415 & 1036 & 399\tabularnewline
\textbf{CART-6} & 0-back & 1892 & 563 & 568 & 118 & 1594 & 815 & 640 & 92 & 626 & 2130 & 385 & 0\tabularnewline
 & 2-back & 1731 & 448 & 678 & 343 & 1818 & 361 & 868 & 153 & 815 & 0 & 2207 & 178\tabularnewline
 & 3-back & 1310 & 266 & 347 & 477 & 751 & 786 & 213 & 650 & 404 & 0 & 63 & 1933\tabularnewline
\hline 
\multirow{1}{1.3cm}{} & Rest & 3366 & 1471 & 1469 & 1053 & 3836 & 1384 & 1222 & 917 & 5595 & 454 & 1043 & 267\tabularnewline
\textbf{CART-10} & 0-back & 1742 & 652 & 549 & 198 & 1448 & 852 & 754 & 87 & 486 & 2470 & 185 & 0\tabularnewline
 & 2-back & 1595 & 559 & 685 & 361 & 1419 & 355 & 1192 & 234 & 408 & 0 & 2782 & 10\tabularnewline
 & 3-back & 1186 & 372 & 348 & 494 & 1036 & 584 & 306 & 474 & 204 & 0 & 63 & 2133\tabularnewline
\hline 
 & Rest & 4795 & 1089 & 1018 & 457 & 3932 & 1759 & 1185 & 483 & 2948 & 1388 & 1217 & 1806\tabularnewline
\textbf{LDA} & 0-back & 2655 & 411 & 67 & 8 & 2269 & 649 & 216 & 7 & 494 & 2063 & 156 & 428\tabularnewline
 & 2-back & 2516 & 137 & 427 & 120 & 2103 & 246 & 518 & 333 & 959 & 130 & 1704 & 407\tabularnewline
 & 3-back & 1722 & 273 & 203 & 202 & 1247 & 446 & 398 & 309 & 408 & 62 & 0 & 1930\tabularnewline
\hline 
\end{tabular}\normalfont

}
\end{table}

\end{document}